\documentclass[journal]{IEEEtran}
\usepackage{blindtext}
\usepackage{latexsym}
\usepackage{amsfonts}
\usepackage{amssymb}
\usepackage{amsbsy}
\usepackage{amsmath}
\usepackage{amsthm}
\usepackage{color}
\usepackage{pgf}
\usepackage{graphicx,lscape,rotating}
\usepackage{balance}
\usepackage[varg]{txfonts}
\usepackage{algorithm,algpseudocode}
\usepackage[noadjust]{cite}
\usepackage[utf8]{inputenc}
\usepackage{bigstrut}
\usepackage{epstopdf}
\usepackage[framed,bw]{mcode}
\usepackage{multirow}
\usepackage{url}
\usepackage{subfig}

\theoremstyle{remark}

\theoremstyle{definition}

\graphicspath{{./images/}}

\begin{document}

\title{Data Aggregation and Packet Bundling of Uplink Small Packets for Monitoring Applications in LTE}

\author{\IEEEauthorblockN{Dong~Min~Kim, Ren\'{e} Brandborg S{\o}rensen, Kashif~Mahmood, Olav~Norvald~{\O}sterb{\o}, Andrea~Zanella and Petar~Popovski}%
\thanks{Part of this work has been supported by the European Research Council (ERC
Consolidator Grant Nr. 648382 WILLOW) within the Horizon 2020 Program. Part of this
work was supported by Innovation Fund Denmark, via the Virtuoso project.}
\thanks{D.~M.~Kim, R.~B.~S{\o}rensen, and P.~Popovski are with the Department of
Electronic Systems, Aalborg University, Denmark (email: \{dmk, rbs,
petarp\}@es.aau.dk).}
\thanks{K.~Mahmood and O.~N.~{\O}sterb{\o} are with Telenor Research, Norway
(email: \{Kashif.Mahmood, olav-norvald.osterbo\}@telenor.com).
}%
\thanks{A.~Zanella is with the Department of Information Engineering, University of
Padova, Italy (email: zanella@dei.unipd.it).}%
}

\maketitle

\begin{abstract}
In cellular massive Machine-Type Communications (MTC), a device can transmit
directly to the base station (BS) or through an aggregator (intermediate node).
While direct device-BS communication has recently been in the focus of 5G/3GPP
research and standardization efforts, the use of aggregators remains a less
explored topic. In this paper we analyze the deployment scenarios in which
aggregators can perform cellular access on behalf of multiple MTC devices. We
study the effect of packet bundling at the aggregator, which alleviates overhead
and resource waste when sending small packets. The aggregators give rise to a
tradeoff between access congestion and resource starvation and we show that packet
bundling can minimize resource starvation, especially for smaller numbers of
aggregators. Under the limitations of the considered model, we investigate the
optimal settings of the network parameters, in terms of number of aggregators and
packet-bundle size. Our results show that, in general, data aggregation can
benefit the uplink massive MTC in LTE, by reducing the signalling overhead.
\end{abstract}


\IEEEpeerreviewmaketitle

\section{Introduction}
\label{sec:introduction}

Machine-type communication (MTC) is growing at an impressive rate, fuelled by the
widespread deployment of Internet of things (IoT) services such as smart metering,
smart grids, e-health, intelligent transport, etc. Predictions are pointing out to
18~billion IoT devices connected to wireless networks in 2022 and
beyond~\cite{EricsonMobiltyReport}. Furthermore, a massive number of machine-type
devices (MTDs) will be connected to the cellular network in regions covered by one
or few Base Stations (BSs). This poses unique challenges to cellular networks that
are tailored for human communication, which is typically downlink-dominated, with
long session times and large packets~\cite{shafiq2013large}. The overhead of channel
access and signalling often represents only a small fraction of the exchanged data
in typical human communication. In contrast, MTC, especially for
monitoring/reporting applications, is usually uplink-dominated, with short session
times and short packets~\cite{shafiq2013large}, the connections are typically set up
for the time needed to transfer a few bytes of payload data, and then teared down.
All this, in the light of the massiveness of MTDs, makes the impact of the overhead
significant. Third generation partnership project (3GPP), recognizing this problem,
standardized three MTC technologies in release 13: extended coverage GSM (EC-GSM)
for 2G networks which improves the legacy network so as to increase coverage
\cite{TS43064}; enhanced MTC (eMTC) for LTE networks; and, finally, the narrow-band
IoT (NB-IoT) technology \cite{TS36211} which can utilize both 4G and 2G spectrum to
provide reliable and secure communication at a low cost.

On an architectural side, the problem of massive signalling overhead can be
alleviated by adopting a 2-stage approach, in which a MTD communicates to the BS via
an intermediate node, here referred to as \emph{aggregator}. The aggregator covers a
spatial region that is (much) smaller than the wide area covered by the BS and
communicates with its associated MTDs via a capillary network, such as WiFi,
Bluetooth or other short-range protocols. While the concept of aggregator assisted
MTC is not new,  to the best of our knowledge there is no work quantifying the
reduction of signalling overhead brought about by aggregators in an LTE scenario,
and taking into account the details of the radio access and radio resource control
procedures. This work tries to fill this gap and details the three key factors which
account for the reduction in the signalling: (1) the aggregation of multiple tiny
flows into a more consistent compound flow makes it possible to keep alive the
connection with the BS, thus reducing the signalling due to multiple session
establishments and tear downs; (2) we enable the aggregator to perform \emph{packet
bundling}, i.e., aggregating multiple small packets from the MTDs into a larger
packet for which the aggregator uses a single access request to the BS, thus
proportionally reducing the transmission overhead; (3) the total number of access
requests to the BS will be reduced, as a single aggregator acts as a proxy for
multiple MTDs. The use of an aggregator is also advantageous for downlink
transmission. Since the aggregator can be placed closer to the MTD than the cellular
BS, the wireless communication distance can be reduced and the reliability of the
downlink transmission can be increased. Furthermore the cost for managing the
massive number of MTDs can be saved because the MTDs do not need to be equipped the
high-cost cellular communication modem.

Traditionally, clustering through aggregators (relays) has been used for coverage
improvement as well as reduced energy consumption of sensor
devices~\cite{abbasi2007survey}. The aggregator-assisted MTC is expected to spread.
For example, many water meters could be
connected via ISM band to aggregators, which are then connected to the BS via
cellular networks. Multiple works have pointed out the negative impact of massive
uplink transmissions on the cellular network
\cite{madueno2016assessment,laya2014random,Biral20151}. Recently, several works
considered serving uplink MTC transmissions by adopting various aggregation schemes
\cite{dawy2017toward,malak2016optimizing,shariatmadari2015data}. In
\cite{dawy2017toward}, the benefits of a clustering access scheme for MTC are
described from both technical and business perspectives at a rather conceptual
level. A multi-level uplink aggregation scheme is presented in
\cite{malak2016optimizing}, where the energy efficiency is analyzed using stochastic
geometry. However, the model in \cite{malak2016optimizing} does not account for the
detailed access reservation procedure used for connection establishment, such that
the effect of reduced signalling overhead is not captured. Instead,
\cite{shariatmadari2015data} proposes a data aggregation scheme in which a gateway
collects the MTC data within a fixed period and then forwards the aggregated data to
an LTE BS. The LTE model used in \cite{shariatmadari2015data} considers a simplified
connection establishment, not taking into account that recently active aggregators
do not need connection establishment, which is one of the key elements in our model
and results. Furthermore, the use of fixed aggregation period introduces fixed,
potentially large, latency for the MTD transmissions.

\begin{figure*}[tb]
\centering
\subfloat[Direct access]{
\includegraphics[width=0.25\linewidth]{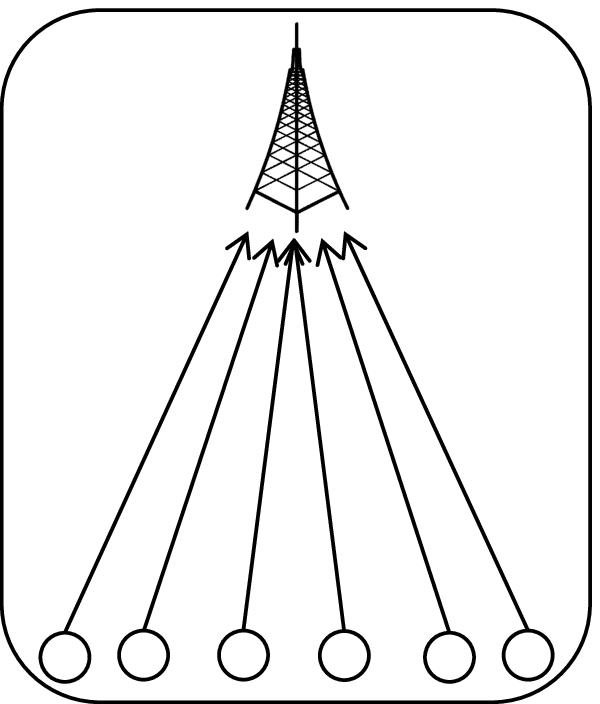}
\label{F:diagram_noagg}}
\subfloat[Single aggregator]{
\includegraphics[width=0.25\linewidth]{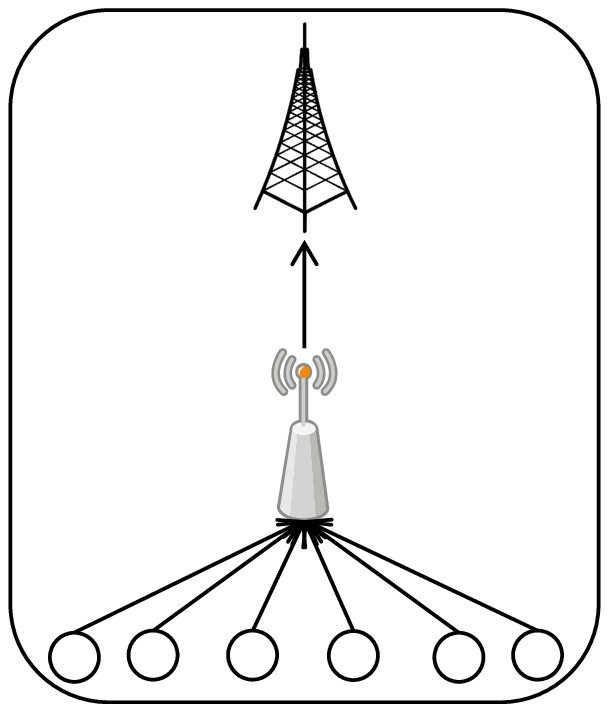}
\label{F:diagram_oneagg}}
\subfloat[Multiple aggregators]{
\includegraphics[width=0.25\linewidth]{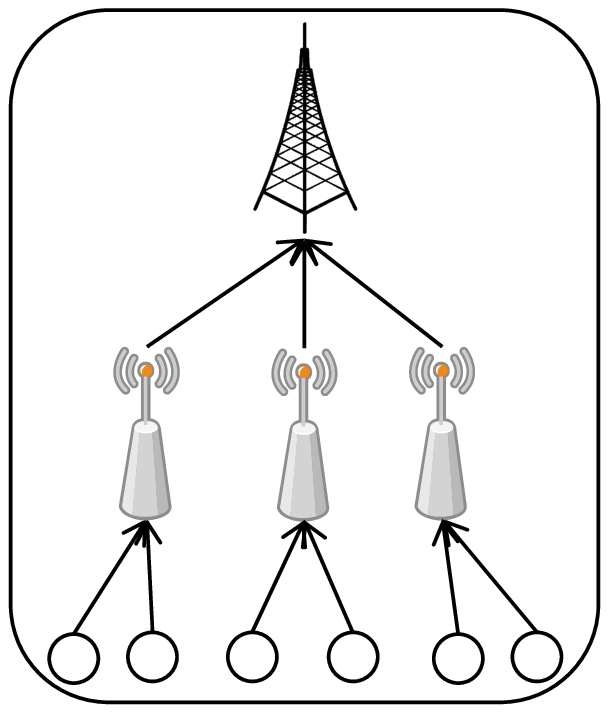}
\label{F:diagram_multiagg}}
\caption{(a) Direct access has maximum diversity, only a single device suffers if a link has low throughput or reliability. (b) A single aggregator means minimal signalling. Transmissions are, however, limited by the throughput and reliability of a single link. (c) Having multiple aggregators is a tradeoff between diversity and minimizing signalling overhead.}
\label{F:access_scheme}
\end{figure*}

In this paper, instead, we consider a more detailed model of the dynamics of the
cellular access process. Specifically, we account for the connection establishment
and release procedures, and for the resource allocation in the physical channels
(PRACH, PDCCH, PDSCH, and PUSCH), which we found to significantly impact on the
performance of the aggregation schemes.

This setup is used to answer a number of research questions:
\begin{itemize}
\item What is the optimal number of aggregators for a certain density of MTDs?
\item How large is the throughput increase brought by the aggregators?
\item Is it possible to aggregate packets without introducing excessive latency?
\end{itemize}

The rest of the paper is organized as follows. We describe the system model for the
cellular network with uplink MTC devices and aggregators in Section~II. In Section~III we
provide the explanation about how the aggregation of massive MTC  works. In Section~IV we
present numerical results. The conclusions are given in Section~V.

\section{System Model}
\label{sec:sys_model}

Assume a cell with $M$ MTDs and $N$ aggregators. In this setup each MTD transmits to
its associated aggregator that, in turn, forwards the transmissions to the BS of the
cell. Therefore, in this paper we investigate the trade-off between diversity gain
and signalling overhead for MTC. In particular, we consider LTE as our case of
study, but we note that the results shown in this paper are applicable to any other
system that requires control signalling and random access to acquire transmission
resources. In our study the capillary connections to the aggregators are idealized,
being free of errors and offering negligible latency, which means that the obtained
results should be treated as upper bounds of entire system performance. In practice,
the capillary networks would introduce errors and certain latency, which depends on
the technology used. In this work, however, we only focus on the aggregating links
between aggregators and BS, which have been assumed to represent a performance
bottleneck.

\begin{figure*}[tb]
    \centering
        \subfloat[Example of network topology. Single LTE-BS with multiple aggregators. Each MTD is associated to the spatially closest aggregator, according to a Voronoi tessellation of the area with respect to the positions of the aggregators.]{
        \includegraphics[width=0.5\linewidth]{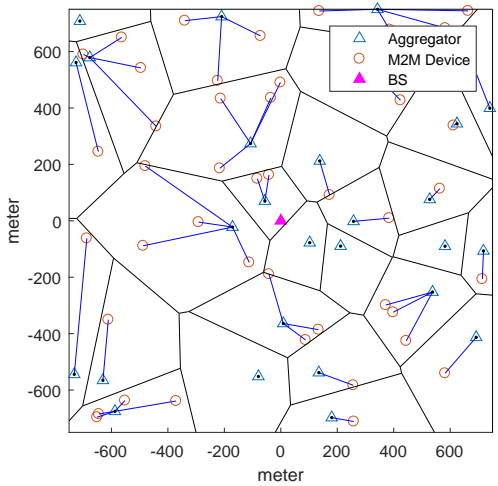}
        \label{F:agg_topology}}
        \subfloat[Example of transmissions in aggregation and packet bundling scheme.]{
        \includegraphics[width=0.5\linewidth]{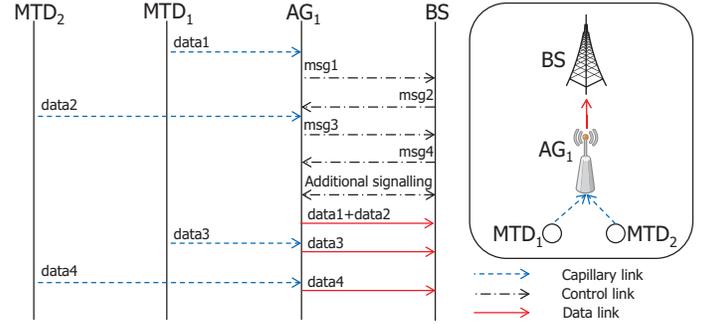}
        \label{F:proposed_scheme}}
	\caption{Considered system model.}
	\label{F:system_model}
\end{figure*}

Different aggregation scenarios are depicted in Fig.~\ref{F:diagram_oneagg}.
Fig.~\ref{F:diagram_noagg} depicts the case without aggregation. In this case, each
MTD connects directly to the BS so that, with ideal channel access, the reliability
is maximized since a failure in one of the links will not impact the rest of
devices. However, in real systems, we have the problem of large signalling overhead
and massive access, which can severely impact the performance of all the MTDs. From
a signalling perspective, and assuming idealized capillary connection to the
aggregator, the optimal solution would be to utilize a single aggregator that
collects the data from all the MTDs, see Fig.~\ref{F:diagram_oneagg}. In this case,
however, the performance of all the MTDs can be compromised by the restrictions in
the throughput of a single link and reliability issues as, e.g., deep fading periods
or blockages of the link, which would impact the service of all MTDs and may lead to
unacceptable degradation of the quality of service (QoS). Furthermore, such a
solution would move the signalling congestion and access problems to the capillary
network, which will make the assumption on idealized capillary invalid. Note that
the capillary network is short-range, which would unrealistically imply that all
MTDs are clustered in a spatial proximity. In any case, as we will see later, the
single aggregator is not found to be the optimal solution even when we neglect these
issues, considering idealized capillary network that offers short-range, low
interference, high reliability, and zero latency communication. Then, the best
solution it to deploy multiple aggregators, as in  Fig.~\ref{F:diagram_multiagg}.
How to determine the optimal number of aggregators as a function of the MTDs density
and packet generation rate is one of the results of our study.

\subsection{Traffic Model}
We assume that time is slotted, and in a given time slot $T_s$, each MTD generates
traffic according to a Poisson distribution with intensity
$\lambda_\text{app}$~[packets/s]. The traffic generated in a time slot is instantaneously
forwarded to the aggregators, based on a nearest neighbour rule. Therefore, the number of
MTDs connected to an aggregator will determine the amount of uplink traffic for that
aggregator.

Additionally, to the aforementioned temporal considerations, our traffic model also
takes into account a spatial component, which is used for the association of MTDs to
aggregators and for link quality evaluation. MTDs and aggregators are assumed to be
uniformly deployed in the cell, following two independent Poisson point processes
with parameters $\lambda_u$ and $\lambda_a$ [nodes/m$^2$], respectively. As
mentioned, each MTD associates to the closest aggregator. Due to the randomness in
the spatial deployment, the number of MTDs served by the different aggregators is
also random, and some aggregators may not be serving any active MTD, as shown in
Fig.~\ref{F:agg_topology}. We define an aggregator as \emph{active} if it serves at
least one MTD. Based on \cite{lee2012coverage}, the density of active aggregators,
${\lambda'_a}$, can then be estimated as follows:
\begin{align}\label{E:active_aggr}
{\lambda'_a} &= {\lambda _a}\left( {1 - \Pr \left[ \mathrm{no\ serving\ MTD} \right]} \right) \nonumber \\
 &= {\lambda _a}\left( {1 - {{\left( {1 + {{3.5}^{ - 1}}{\lambda _u}/{\lambda _a}} \right)}^{ - 3.5}}} \right).
\end{align}
The average number of devices per active aggregator, in turn, is given by
$\lambda_u/\lambda'_a$. Therefore, the packet arrival process at an active
aggregator is the compound of a random number of independent Poisson generation
processes and, hence, is still Poisson, with rate equal to
$\lambda_u/\lambda'_a\lambda_\text{app}$, which accounts for the impact of spatial
randomness of the nodes on the traffic. However, this traffic model does not
consider temporal or spatial correlations in the packet generation processes, which
may be found in some cases, nor does it consider exception events with irregular
behaviour. Nonetheless, it allows us to get insights into the performance of the
system in a stationary scenario.

\subsection{System Parameters}
We consider a bandwidth of 1.4~MHz for the LTE cell, which corresponds to a Physical
Random Access Channel (PRACH) of 6 resource blocks (RB) in frequency division (FD)
and 1~ms in time division (TD). Every 10 subframes there is a random access
opportunity (RAO). Retransmissions are allowed 4 times and the maximum number of new
random access attempts per payload is 10. The RRC idle timeout is 100~ms.

\subsection{Performance Metrics}
The QoS of the aggregation scheme will be evaluated with regard to latency, outage and
throughput. Furthermore the optimal throughput and the associated number of aggregators
will be found. More specifically, we adopt the following definitions for the different
performance indexes:
\begin{itemize}
\item \emph{Latency} - the period between packet generation at the MTD and
    successful delivery to the BS.
\item \emph{Outage} - the fraction of generated packets that are not successfully
    received within the simulation time.
\item \emph{Throughput} - the total amount of successfully transmitted data in bits divided by the sum of the latencies of each successful transmission.
Note that, according to this definition, the throughput is a measure of the average bitrate experienced by each packet transmission and, hence, it is not limited by the traffic generation rate.
\item \emph{Optimal throughput }- the maximum throughput that can be obtained for
    a given number of MTDs.
\item \emph{Optimal number of aggregators} - the number of aggregators needed to obtain the
    optimal throughput.
\end{itemize}

\section{Aggregation and Packet Bundling Scheme}
\label{sec:delay_model} In this section, we describe the proposed aggregation and
packet bundling scheme in more detail.

As shown in Fig.~\ref{F:proposed_scheme}, data generated by the MTDs is first
delivered to the aggregator, using another technology (which we assume orthogonal to
LTE) that offers ideal capillary connection, as already described. The aggregator
forwards this data to the BS using the LTE uplink channel. If the aggregator is not
already connected to the BS, it needs to perform a random access procedure in order
to acquire the transmission resources. To this end, the aggregator first transmits a
random preamble (msg1) in a PRACH slot. The BS's reply (msg2) indicates where to
send the connection request message (msg3). If the connection request is not
accepted by the BS, then the aggregator repeats the random access procedure.
Otherwise, the BS sends a contention resolution message (msg4) and, with some
additional signalling, the connection is established and the aggregator gets
assigned exclusive access to the required resource Blocks (RBs) in the uplink
channel (PUSCH), which can then be used to send data. The aggregator retains a
dedicated resource in PUCCH for transmission of new scheduling requests (SRs) until
it is disconnected by the Radio Resource Control (RRC) after a sufficiently long
idle period. Bringing this additional signalling to LTE in the PUCCH can be
justified by simulations in \cite{TR36822}, which found the PUCCH utilization to be
very low. In \cite{TR36822} less than 0.1 \% scheduling request opportunities are
used for 10ms scheduling request period.

The packet bundling aims at improving the system efficiency by making a better use
of the resources allocated to the node in an opportunistic manner. At the time of SR
for transmission an aggregator will bundle maximum $B$ packets in its transmission
buffer to the single packet, which first triggered the SR. This mechanism is mainly
triggered during the connection establishment procedure, when packets that are
received by the aggregator while performing the access procedure are bundled, as for
\texttt{data1} and \texttt{data2} in Fig.~\ref{F:proposed_scheme}. Clearly, the
resource request will be dimensioned on the size of the bundled packet
(\texttt{data1+data2}), rather than on that of the packet that has started the
process (\texttt{data1}). However, to avoid excessive resource requests in case of
massive packets arrivals during the connection establishment phase, the maximum
number of packets that can be bundled together is limited to $B$, and the excess
packets are simply buffered and sent after the connection is established. In the
connected state, the aggregator immediately sends the received packets as for
\texttt{data3} and \texttt{data4} in the figure. However, if multiple packets arrive
at the aggregator during an ongoing transmission, they are bundled together.

\section{Simulation of Data Aggregation and Packet Bundling in LTE}
\label{sec:lte_simulation} In this section we describe the LTE simulator and the
numerical results.

The term User Equipment (UE) is used in the following to describe devices connected
to the BS, i.e., the aggregators when $N>0$, and the MTDs when $N=0$, since in
absence of aggregators the MTDs are connected directly to the BS. We consider the
latter case as a benchmark for the aggregation scheme.

\begin{table}[tb]
\centering
\caption{Simulation parameters}
\label{T:trafficparams}
\begin{tabular}{|l|l|l|}
\noalign{\hrule height 1pt}
Metric                    	& Designation 	& Value		\\ \noalign{\hrule height 1pt}
\textbf{Traffic distribution parameters} & & \\ \noalign{\hrule height 0.75pt}
Cell radius                 & -           & 1000 m                  \\ \hline
Number of MTDs       & $M$           & Variable                \\ \hline
Number of aggregators       & $N$           & Variable                \\ \hline
Packet arrival rate per MTD& $\lambda_{\rm app} $          &  Variable              \\ \hline
Packet size         		   & -             & 100 bytes                 \\ \noalign{\hrule height 1pt}
\textbf{PHY parameters} & & \\ \noalign{\hrule height 0.75pt}
EARFCN	 				& -				& DL:5900						\\ \hline
Downlink Tx power    	& -            	& 30 dBm              \\ \hline
Uplink Tx power    		& -            	& 23 dBm              \\ \hline
\textbf{MAC parameters} & & \\ \noalign{\hrule height 0.75pt}
Number of RACH preambles	 	 				& -				& 54						\\ \hline
Backoff time 	 				& -				& 20 subframes			\\ \hline
Maximum RACH retransmissions   	& $K$            	& 10                      	\\ \hline
Random access opportunities  	& -            	& 1 every 10 subframes 	\\ \hline
Fragmentation threshold			& -				& 6 RBs\footnotemark		\\ \noalign{\hrule height 1pt}
\textbf{System parameters} & & \\ \noalign{\hrule height 0.75pt}
Number of RBs per RACH slot	&  -           	& 6       	\\ \hline
Simulation length           	& $T_s$            	& 60 s      	\\ \hline
Maximum data retransmissions	& $L$            	& 1        	\\ \hline
Maximum number of bundled packets			& $B$				& Variable        	\\ \hline
Processing time 				& -           	& 3 ms    	\\ \noalign{\hrule height 1pt}
\end{tabular}
\end{table}
\footnotetext{If a packet cannot be transmitted in a single subframe, it is
fragmented.}

\subsection{Simulation}

The simulation has been developed in MATLAB, accounting for all the details of the
LTE channel access procedure.  More specifically, each simulation run consists in
the following six steps:
\begin{enumerate}
\item Configuration of the simulation parameters.
\item Random placement of the MTDs and the aggregators in the cell.
\item Random generation of the packet arrivals for all MTDs.
\item Event-based simulation of the channel access procedure and packet
    transmission according to the LTE specifications.
\item Processing of the results and averaging over multiple repetitions of steps
    2-4.
\item Post processing of the results and visualization.
\end{enumerate}
In a two-dimensional region, MTDs and Aggregators are randomly distributed within
the single cell. Therefore, each aggregator serves a different number of MTDs and
the random distance between the aggregator and the base station affects the
performance of the wireless transmission. We have HARQ, but the transmission power
is fixed and we have thus disabled adaptive modulation as well.
Table~\ref{T:trafficparams} collects the setting of the simulation parameters.

As mentioned, the simulation is event-based. Each event corresponds to the
transmission of a \emph{message}, which can be part of the signalling, RACH, or data
transfer procedure. All messages are listed in a virtual queue and ordered according
to the transmission instant.  At each simulation step, the next message in the
virtual queue is fetched and the corresponding event is simulated. If the event is a
new packet arrival to a UE that is not connected to the BS, then the RACH procedure
is simulated (assuming 6 signalling messages after msg4 to allocate resources in the
PUSCH to the UE). If the procedure is successfully completed, the UE switches to the
connected RRC state and starts a \emph{connection timer} which is renewed upon any
successful transmission. If the connection timer expires, the UE releases the
resources and tears down its connection to the BS. If the event is a packet arrival
at a UE that is already connected to the BS, uplink resources are requested on the
PUCCH and the message will be sent to the BS using the granted PUSCH resources, if
any are available. The SR on the PUCCH is implicitly handled in the simulation and
the grant may take place soonest possible, at the next subframe. In case of
transmission errors due to collisions during the RACH procedure, channel
fluctuations, or time-outs due to resource starvation, retransmissions can be
inserted in the virtual queue, as for the simulated LTE protocols.

\subsection{Numerical results}

\begin{figure*}[tb]
\centering
\subfloat[Throughput]{
\includegraphics[width=0.33\linewidth]{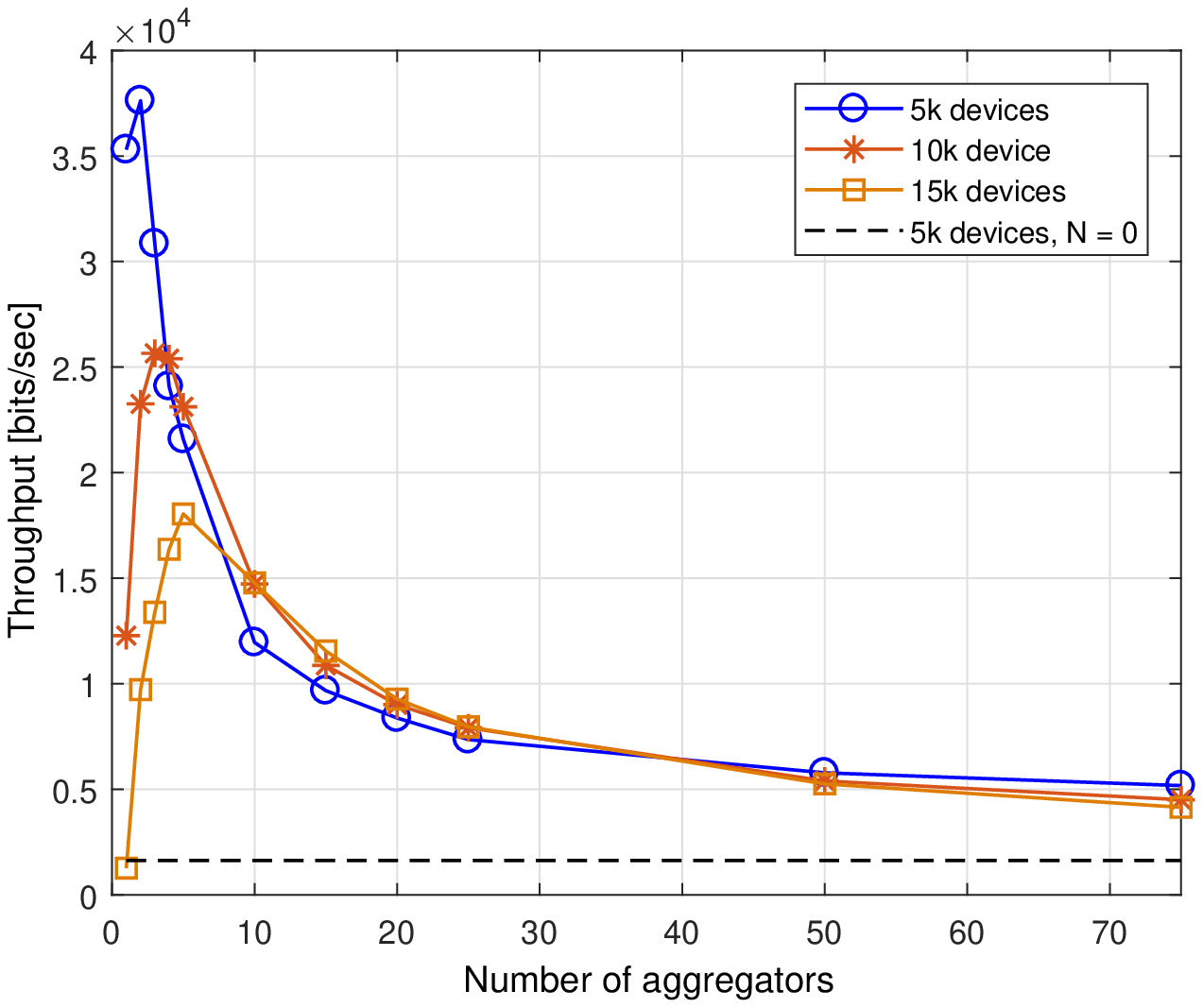}
\label{F:Throughput}}
\subfloat[Latency]{
\includegraphics[width=0.33\linewidth]{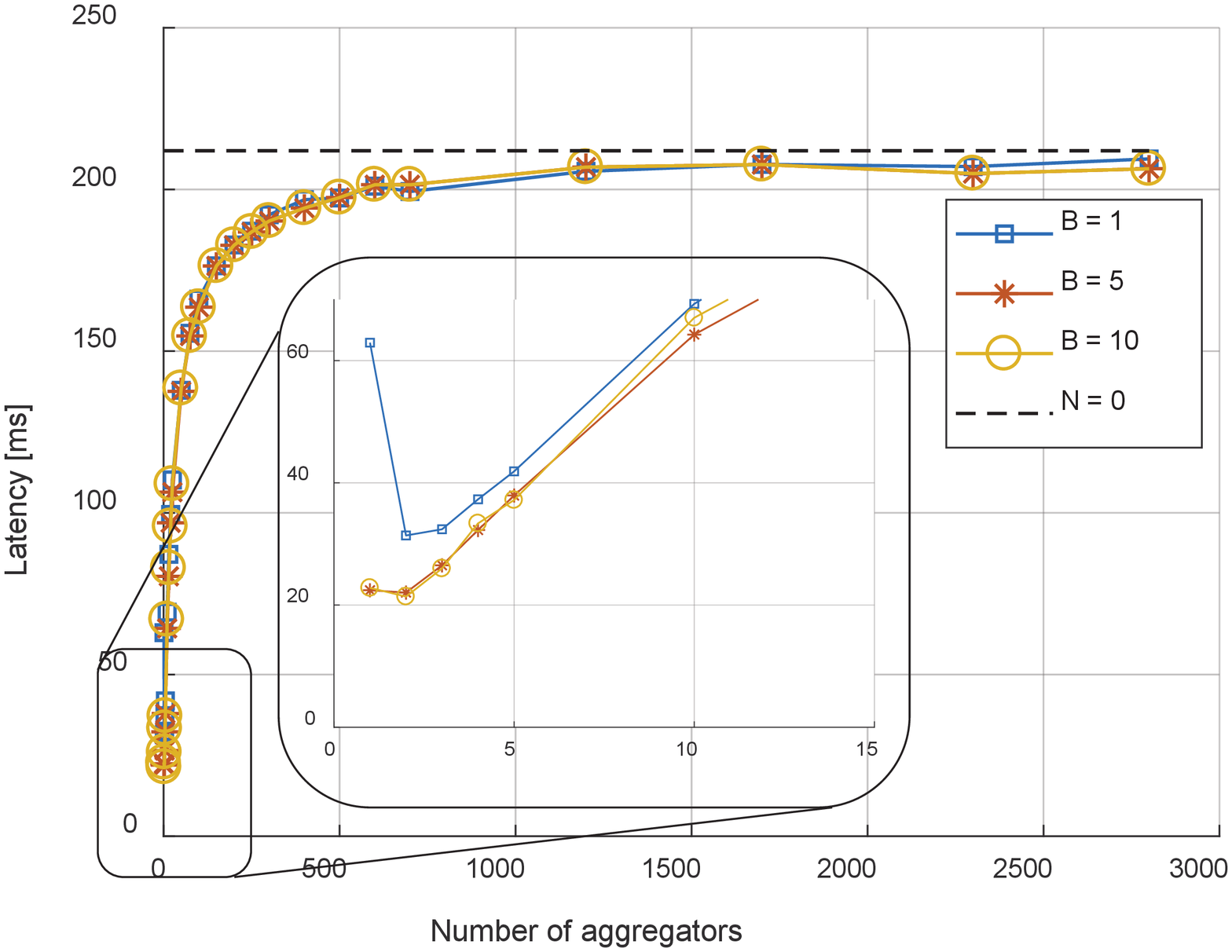}
\label{F:Latency}}
\subfloat[Outage]{
\includegraphics[width=0.33\linewidth]{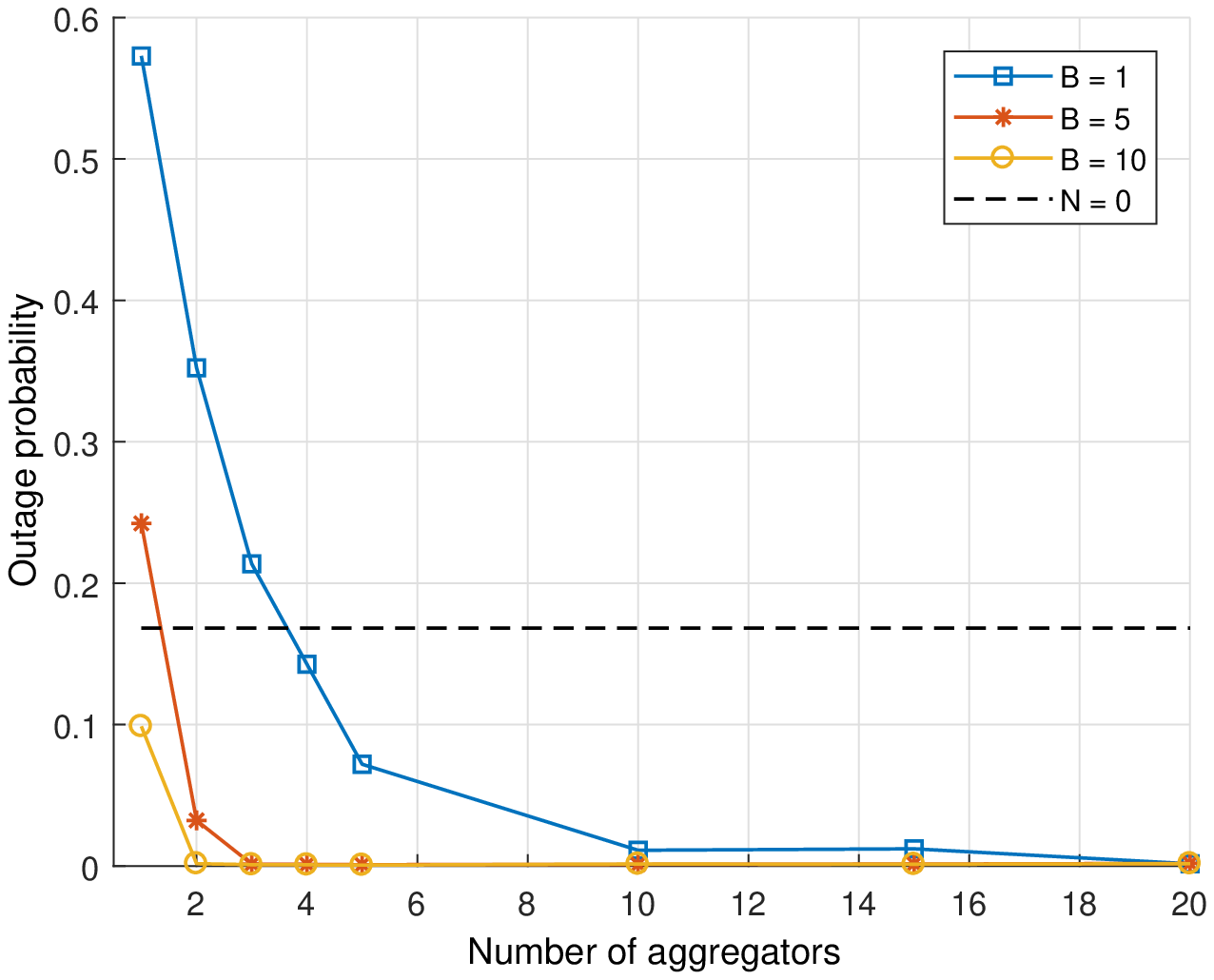}
\label{F:PacketBundlingOutage}}
\caption{(a) Throughput as a function of the number of aggregators, $\lambda_{\rm app}= 1$~[packet/min] and $B = 10$.  (b) Latency as a function of the number of aggregators, with $M = 5000$ and $\lambda_{\rm app}= 1$~[packet/min]. (c) Effects of packet bundling on the outage as a function of the number of aggregators, with $M = 5000$ and $\lambda_{\rm app}= 3$~[packet/min].}
\label{F:performFigure}
\end{figure*}

The throughput of the system is plotted in Fig.~\ref{F:Throughput} as a function of
the number of aggregators, when varying the number of MTDs. We can observe that the
throughput is low when the number of aggregators is very low (reduced spatial
diversity) or very large (channel access contention). Note that, despite the ideal
capillary network, having a single aggregator is not optimal. The reason is that the
number of capillary arrivals is larger than the service rate of the LTE link.
Hence, there exists an optimal number of aggregators, larger than one, that
maximizes the throughput of the system.

Fig.~\ref{F:PacketBundlingOutage} reports the outage probability when varying the
number of aggregators, for a population of $M=5000$ MTDs with a packet generation
rate of $\lambda_{\rm app}=3$ packets per minute. The different curves have been
obtained by changing the number $B$ of packets that can be bundled together during
the RACH procedure. Furthermore, the curve without aggregators ($N=0$) is added as
benchmark. We observe that the outage rapidly raises when the number of aggregators
drops below a certain threshold, because the compound arrival-rate at an aggregator
exceeds its link capacity. However, packet bundling enhances the capacity of the
aggregators, shifting to the left the point at which the outage increases. This
capacity gain is due to the more efficient use of the RBs assigned to the
aggregator. The bundling limiter, $B$, can be replaced with a size limit for the
bundled packet to accommodate realistic machine type traffic with varying packet
sizes. As the packet sizes are fixed in this work, $B$ is a good indicator of what
would be the behaviour of such a limiter.

The latency can be found in Fig.~\ref{F:Latency} where the latency is seen to be
high for very low numbers of aggregators. This is due to the capacity of a single
aggregator being limited. This may be alleviated to some degree by using a larger
value $B$ for the maximum packets bundled. The latency also grows as a larger number
of aggregators are competing going towards the latency of the benchmark case. Thus
an optimal operational point can be found.

\begin{figure}[tb]
\centering
\includegraphics[width=0.9\columnwidth]{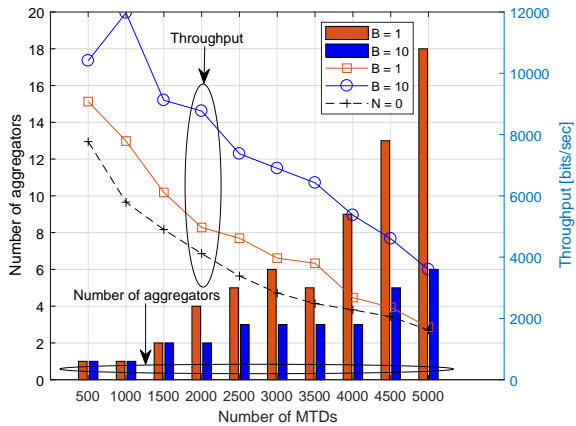}
\caption{Optimal throughput and the associated number of aggregators as a function of the number of MTDs for $B=1$ (red) and $B=10$ (blue), $\lambda_{\rm app}= 1$~[packet/min].}
\label{F:optaggs}
\end{figure}

Fig.~\ref{F:optaggs} reports the optimal throughput (lines) and the corresponding
optimal number of aggregators (bars) as a function of the number of MTDs, with
($B=10$) and without ($B=1$) packet bundling. In addition, the figure also reports
the throughput (dashed line) for the baseline case with no aggregators ($N =0$). We
can see that, with packet bundling, the  optimal number of aggregators grows more
slowly with the number of MTDs compared to the case without packet bundling, thus
confirming the capacity gain previously observed. It can also be noticed that the
achievable optimal throughput decreases almost linearly with the number of MTDs,
which agrees with the results in Fig.~\ref{F:Throughput}.

\begin{figure}[tb]
\centering
\includegraphics[width=0.9\columnwidth]{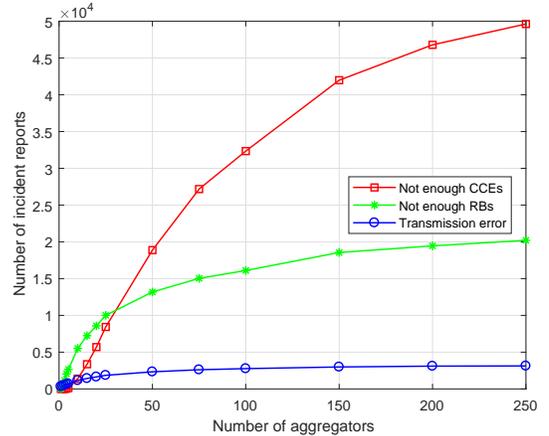}
\caption{Incident reports for transmission delays for $M = 5000$, $B = 10$ and $\lambda_{\rm app}= 1$. Starvation of resources in either PDCCH or PUSCH can lead to message expiration as transmissions can not occur if either channel lacks resources.}
\label{F:Bottlenecks}
\end{figure}

The incident reports of Fig.~\ref{F:Bottlenecks} show how many times delays were
reported due to each of the factors: transmission error, lack of control channel
elements (CCEs) and lack of RBs. This plot indicates that the PDCCH has a relatively
larger effect on the latency of the system as the number of aggregators grow, while
the PUSCH has a larger impact on the performance for less aggregators.

\section{Conclusions and Future Work} \label{sec:conclusions}

In this paper, we studied the effects of an aggregation scheme to sustain the
uplink MTDs traffic in a LTE scenario. Our LTE model accounted for the details of
the radio access and radio resource control procedures that, depending on the
arrival rate and intensity of the traffic flows, are fundamental to fully capture
the impact of traffic aggregation. We also proposed a packet bundling mechanism,
which was found to further improve the capacity of the system in terms of the
supported number of devices per aggregator and throughput of each aggregator. We
evaluated the number of aggregators needed to optimize the throughput of each
aggregator  with and without packet bundling, under the assumption that the
connection between MTDs and aggregators is obtained by means of ideal (zero latency,
zero outage, infinite capacity)  capillary networks.

Our results clearly show that, when evaluating schemes that involve aggregators, it
is important to take into account the details of the cellular technology, such as
RRC in LTE, as aggregation has a large impact on the number of access attempts and,
in turn, on the signalling overhead seen at the BS. Furthermore, packet bundling
turns out to be a promising strategy for aggregation in capillary cells. The
evaluated packet bundling mechanism added no obligatory queueing delay at the
aggregator whilst enhancing the throughput of each aggregator. From another
perspective, we can state that packet bundling makes it possible to lower the number
of aggregators required to optimally serve a given MTD density, thus reducing the
overall cost of the infrastructure that includes aggregators.

Possible research directions include a thorough end-to-end performance analysis by
specifying a communication model for the capillary networks. Furthermore, it is
relevant to optimize resource management and packet bundling in the presence of
heterogeneous MTDs with more realistic traffic-generation models. As massive MTC is
one of the important use case for 5G cellular networks, the aggregator is highly
effective method to bear massive MTC traffic. Design of reliable signaling schemes
and low-overhead access protocol for the aggregator is essential to address the
requirements for massive MTC application in 5G.

\ifCLASSOPTIONcaptionsoff
  \newpage
\fi



\end{document}